\documentclass[final,aps,pra,12pt]{revtex4}
\usepackage{graphicx}
\usepackage{amsmath}
\def\bea{\begin{eqnarray}}
\def\eea{\end{eqnarray}}

\def\ba{\begin{array}}
\def\ea{\end{array}}
\def\({\left(}
\def\){\right)}
\def\[{\left[}
\def\]{\right]}

\def\sech{{\rm sech\,}}

\def\sn{{\rm sn\,}}
\def\cn{{\rm cn\,}}

\def\sin{{\rm sin\,}}
\def\cos{{\rm cos\,}}
\begin{document}
\title{Controlling nonautonomous matter waves in ``smart" transient trap variations}
\author{S. Sree Ranjani} 
\email{s.sreeranjani@ifheindia.org}
\affiliation{Department of Physics, Faculty of Science and Technology (IcfaiTech), ICFAI Foundation for Higher Education, Hyderabad, India.}
\author{Tangirala Shreecharan} 
\email{shreecharan@ifheindia.org}
\affiliation{Department of Physics, Faculty of Science and Technology (IcfaiTech), ICFAI Foundation for Higher Education, Hyderabad, India.}
\author{Thokala Soloman Raju}
\email{solomanr.thokala@gmail.com}
\affiliation{Department of Physics, National Institute of Science and Technology, Palur Hills, Brahmapur, Odisha 761008, India.}         
\begin{abstract}
In this paper, we study the controllable behavior of nonautonomous matter waves in different ``smart" transient trap variations in the context of the cigar-shaped Bose-Einstein condensates. By utilizing a self-similarity transformation we reduce the nonautonomous Gross-Pitaevskii (GP) equation to the elliptic equation that admits soliton solutions. This procedure leads to a consistency equation which is in the form of Riccati equation. The connection between the  Riccati and the linear Schr\"odinger equation, through the Cole-Hopf transformation, is exploited profitably here to introduce temporal trap variations. For our study, we explore the possibility of using  one dimensional exactly solvable (ES) potentials and their newly constructed rational extensions, as functions of time to introduce interesting temporal trap modulations. The fact that the regular potentials and their rational extensions being structurally different, leads to different temporal modulations. It is exhibited that the soliton behavior with respect to compression in both these cases is quite different.       
\end{abstract}
\maketitle

\section{Introduction}

Observation of Bose–Einstein condensates (BECs) in dilute vapors of alkali atoms is a fascinating phenomenon that has opened up a plethora of technological applications \textbf{references needed}. Indeed, BECs \cite{bose,MOM,scin} can be confined to
one and two dimensions by applying suitable traps \cite{bloch,kad,lecoq}. Last few years have seen a flurry of investigations demonstrating variety of atom traps. This has lead to interesting configurations of BECs where a two-dimensional configuration refers to BEC on a chip \cite{pas}. Similarly in a quasi-one-dimensional scenario \cite{billy,ying,fat} one obtains a cigar cigar-shaped BECs. In the latter case, bright \cite{kha,car2,lia,cor,kon,boris} and dark \cite{stef,ADJ,cal} solitons have been observed
in repulsive and attractive coupling regimes, respectively. Here the bright and dark solitons
correspond to rarefactions and density lumps 
compared to the background respectively. It is intriguing to note that the corresponding mean field
Gross–Pitaevskii (GP) equation is nothing but the familiar nonlinear
Schr\"odinger equation (NLSE) in quasi-one dimensions \cite{carr1}, that admits solitons as solutions due to the delicate balance between dispersion and nonlinearity.

The coherent control of solitons in cigar-shaped BECs can be achieved by controlling the variation in the parameters like: scattering length and trap frequencies. The temporal variation of scattering length due to Feshbach resonance has been extensively studied in the literature \cite{Xi,zhang,jing}. The variations in the transverse trap frequency leads to a modulation of the nonlinearity with time. This has been systematically studied in the context of Faraday modes \cite{Nich,atre}. Recently, Serkin et al have revealed the  main features of nonautonomous matter-wave solitons near the Feshbach resonance in a one-dimensional Bose-Einstein condensate confined by a harmonic potential with a varying-in-time longitudinal trapping frequency \cite{serkin1,serkin2,serkin3}. In the context of optical fiber, nonautonomous solitons and their dynamical behavior has been studied extensively \cite{kruglov,soloman1,soloman2}. 

So far in the literature, few studies have shown that the dynamics of the solitons in BECs can be controlled by modulating the trap frequencies in time (refe: sree solomon etc). In the present study, we have elucidated the effect of temporal modulations of the trap, on the ensuing soliton solutions. The modulations are introduced through a consistency conditions obtained in the process of mapping nonautonomous GPE to the elliptic equation \cite{atre}. The consistency condition is in the form of the well-known Riccati equation and can be transformed into a second order differential equation of the Strum Liouville-type using the Cole-Hopf transformation.
 
One of the well studied Strum Liouville-type differential equations is the Schr\"odinger equation. The eigenvalues and eigenfunctions of the latter for one dimensional potentials, can be exploited in the context of NLSE. In our case, we consider these potentials as functions of time which can be used to provide a wide range of temporal trap modulations. In this paper we explore these connections and more pertinently, exploit the Cole-Hopf transformation and its intimate connection with the Ricatti and the Schr\"odinger equations. The potential functions we are going to consider are ES potentials as functions of time. 

Recently the number of ES potentials in quantum mechanics has seen a large growth with the construction of rational extensions of the regular ES potentials. By rational extensions we mean that the regular ES potentials has additional rational terms. Interestingly the regular and the corresponding
rational potentials are isospectral to each other. It is worth mentionaing that the regular potentials have the classical orthogonal
polynomials as their eigenfunctions while the rational potentials have exceptional orthogonal polynomials. To the best of the knowledge of the authors, these rational potentials as ``smart" transient trap variations have been exploited for the first time, in the context of BECs.

\section{Solutions of the NLSE}
   
A cigar shaped BEC is obtained by confining a dilute Bose gas in a cylindrical harmonic trap given by
\begin{equation}
V= \frac{1}{2} m^{\prime}\omega^2_{\bot}(x^2+y^2) +
\frac{1}{2} m^{\prime}\omega^2_0(t)z^2 ,   
\end{equation}
with the time dependent confinement in the $z$ direction. The mean field GP equation reduces to 
the quasi-one dimensional nonlinear Schr\"odinger equation (NLSE)  using a Guassian trial wave function \cite{atre,jack,sal} in a tight confinement regime, $(\omega_{\perp}>>\omega_z)$, applied in the $z$ direction  
\begin{equation}
\imath\partial_{t}\psi=-\frac{1}{2}\partial_{zz}\psi+\gamma(t)|\psi|^{2}\psi+\frac{1}{2}M(t)z^2\psi. \label{e1}
\end{equation}
Where $\gamma(t)= 2a_s(t)N\hbar\omega_{r}$,
$M(t)=\omega_0^2(t)/\omega^2_{\perp}$, $a_{\perp}=\hbar/m^{\prime}\omega_{\perp}$, $\omega_{r}$ is the
transverse harmonic oscillator frequency, $a_s$ is the scattering length and $N$ is the total number of particles. The trap potential becomes confining if $M(t)>0$ and repulsive if $M(t)<0$. 

The following ansatz is used to obtain the exact solutions of the NLSE
\begin{equation}
\psi(z,t)=\sqrt{A(t)}F\{A(t)[z-\ell(t)]\}\exp[\imath\Phi(z,t)],
\label{e2}
\end{equation}
where
\begin{equation}
\Phi(z,t)= a(t)-\frac{1}{2}c(t)z^2 \label{e3z}
\end{equation}
with 
\begin{equation}
a(t)=a_0+\frac{\lambda -1}{2}\int_0^t
A^2(t^{\prime})dt^{\prime}
\end{equation}
and $a_0$ being a constant. In the above equations $a(t)$ and $c(t)$ represent the phase offset and the phase front curvature of the phase profile respectively. $A(t)$ and $\ell(t)$ give the amplitude and the location of the center of mass of the soliton respectively. Substituting the ansatz Eq. (\ref{e2})  of in Eq. (\ref{e1}) leads to the elliptic differential equation 
\begin{equation}
F^{\prime\prime}(T)-\lambda F(T) +2\kappa F^3(T)=0  \label{e3}
\end{equation}
with $\kappa=-\gamma_0/A_0$, $T=A(t)(z-\ell(t))$ and the prime denotes differentiation with respect to $T$. Jacobi elliptic functions \cite{han}, like $\sn(T,m), \cn(T,m)$ etc with elliptic modulus $m$ appear as solutions to the above differential equation. Furthermore one obtains the following consistency conditions for the control parameters
\begin{equation}
A(t)=A_0\exp \left(\int_0^t c(t^\prime)dt^{\prime}\right),\quad
\gamma(t)=\gamma_0 \frac{A(t)}{A_0}, \label{e6a}
\end{equation}
\begin{equation}
\frac{d\ell(t)}{dt} +c(t)\ell(t)=0,   \label{e6b}
\end{equation}
\begin{equation}
\frac{dc(t)}{dt}-c^2(t)=M(t),  \label{e4}
\end{equation}
with $A_0$ and $\gamma_0$ being constants.

In this study we look at the dynamics of the bright solition which arises in the attractive nonlinear regime ($\gamma_0<0$) and the limit $m \rightarrow 1$. Under these conditions 
\begin{equation}
F(T)= \cn(T/\tau_0,m) \rightarrow  \sech(T/\tau_0).
\end{equation}
Thus the bright soliton solution is
\begin{equation}\label{bsol}
\psi(z,t)=\sqrt{A(t)}\sech(T/\tau_0)\exp[\imath a(t)-\frac{\imath}{2}c(t)z^2],
\end{equation}
with $\tau_0^2=-A_0  /\gamma_0$ and $\lambda = -1/\tau_0^2$.

\section{Temporal modulations of the trap}

In the present section we present a method that has been developed and used in Ref. \cite{} to generate temporal trap modulations leading to interesting soliton dynamics. Our starting point is the equation governing the phase front curvature $c(t)$ given by equation (\ref{e4}). It can be noticed that this is the Riccati equation. Using the  Cole-Hopf transformation:
\begin{equation}
c(t)=-\frac{d}{dt}\ln[\phi(t)], \label{e5}   
\end{equation}
in Eq.\eqref{e4} we obtain
\begin{equation}
-\phi^{\prime\prime}(t)-M(t)\phi(t)=0. \label{e6}
\end{equation}
By choosing the ratio of the trap frequencies as 
\begin{equation}\label{trap-pot}
M(t)= M_0+V(t)
\end{equation}
Eq. (\ref{e6}) becomes the Schr\"{o}dinger equation in $t$. This enables us to introduce a large variety of temporal modulations of the trap through the function $V(t)$. This is accomplished by choosing $V(t)$ to be of the same functional form as the one dimensional potential models that one is familiar in quantum mechanics. We therefore have the advantage of knowing the exact eigenfunctions which in turn allow us to determone $c(t)$ via equation Eq. (\ref{e5}). In addition to the aforementioned advantage, Eq. \eqref{e5} allows us to determine the control parameters via equations \eqref{e6a} and \eqref{e6b} in terms of $\phi(t)$ as
\begin{equation}\label{e7}
A(t)=A_0\frac{\phi(0)}{\phi(t)} \quad  \gamma(t)=
\gamma_0 \frac{\phi(0)}{\phi(t)} \quad \ell(t)=\ell_0\frac{\phi(t)}{\phi(0)},
\end{equation}
where $\ell_0$ is a constant. These control parameters are dependent on time and can be exploited to control the soliton dynamics. 

\section{Soliton dynamics}

In this section, we discuss the spatio-temporal dynamics of BECs for different trap modulations introduced using the regular potentials and their rational extensions. This study concentrates on how this rationalization affects the soliton dynamics. The modulations considered are of the form of the regular oscillator, Scarf-I potentials and their rational extensions.

\begin{figure}
\centering
\includegraphics[width=0.9\linewidth]{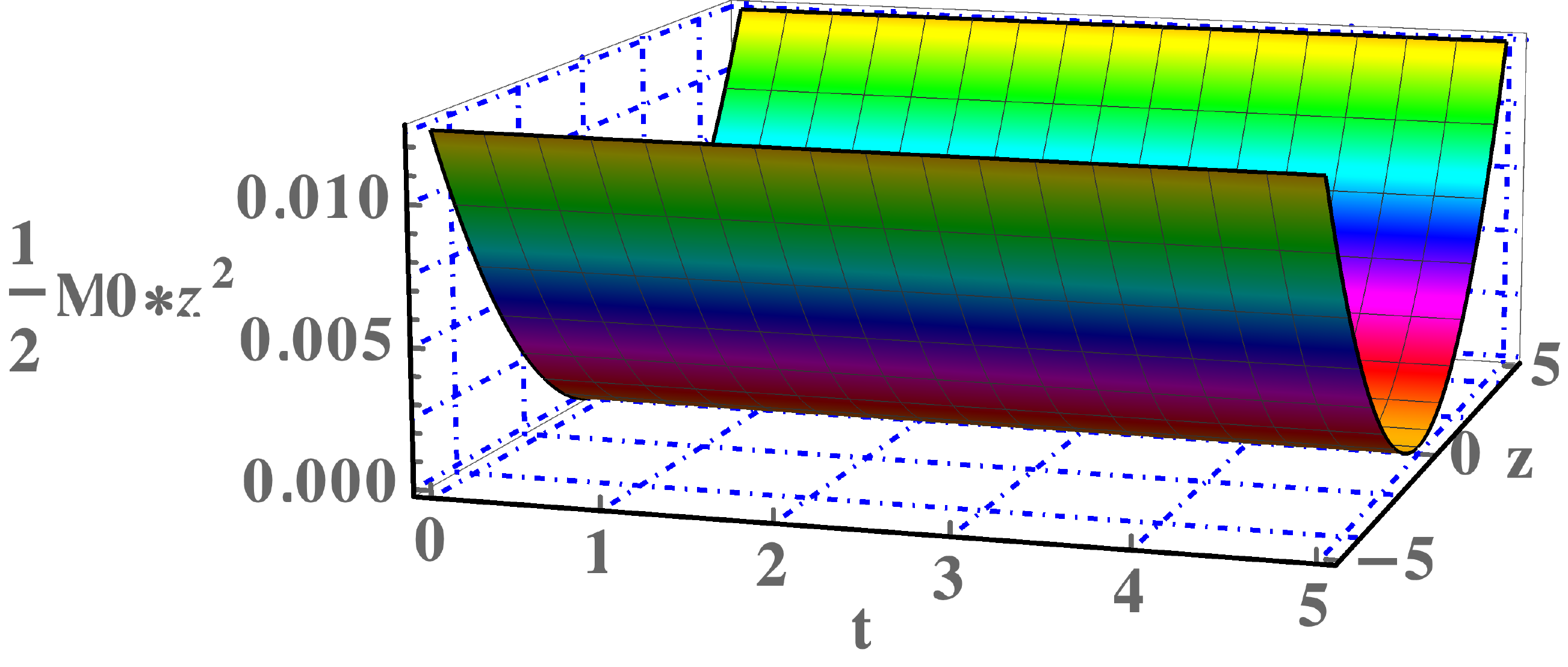}
\caption{Unmodulated trap $M_0 = 0.001$.}
\label{M0.001}
\end{figure}

\begin{figure}
\centering
  \includegraphics[width=0.9\linewidth]{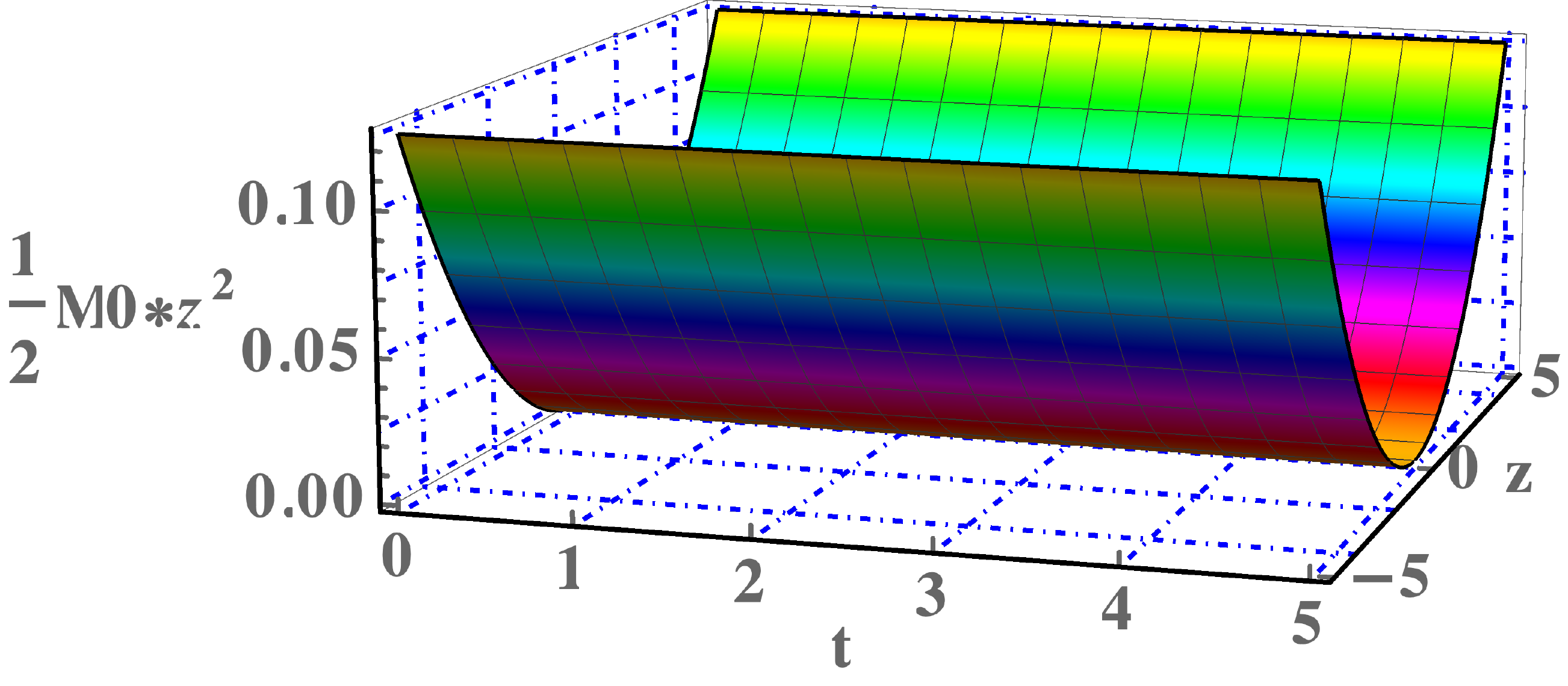}
  \caption{Unmodulated trap $M_0 = 0.01$.}
  \label{M0.01}
\end{figure}

We first present the trap and the spatio-temporal dynamics  of the bright soliton, where the ratio of the trap frequencies $M(t)$ is taken as a constant. We refer to  this as an unmodulated trap. $M(t)$ is a constant implies $V(t)=0$ in Eq. \eqref{trap-pot}, which now takes the form of the free particle Schro\"odinger equation in time,  with one solution  being $\phi(t)= \exp(\sqrt{M_0}t)$.  The trap potential in Eq.\eqref{e1} now becomes  $\frac{1}{2}M_0z^2$, which is a confining parabolic trap as shown in Figs. (\ref{M0.001}) and (\ref{M0.01}) for   $M_0=0.001$ and $M_0=0.1$ respectively. As  can be seen from these figures, the parameter $M_0$ controls the height of the trap. 

\begin{figure}
  \centering
  \includegraphics[width=0.9\linewidth]{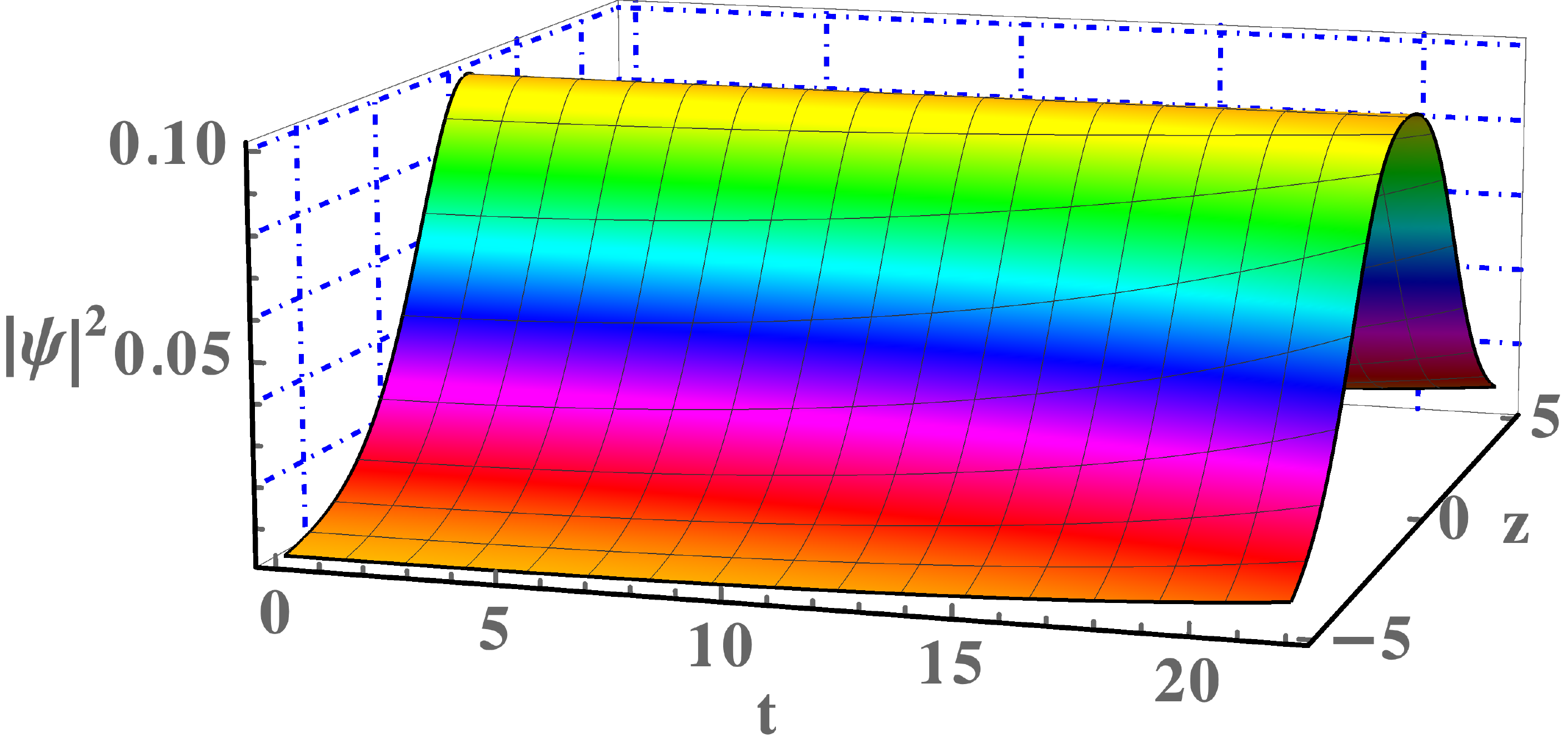}
  \caption{Soliton solution for the unmodulated trap with $M_0=0.001$}
  \label{unmod0.001}
\end{figure}

\begin{figure}
  \includegraphics[width=0.9\linewidth]{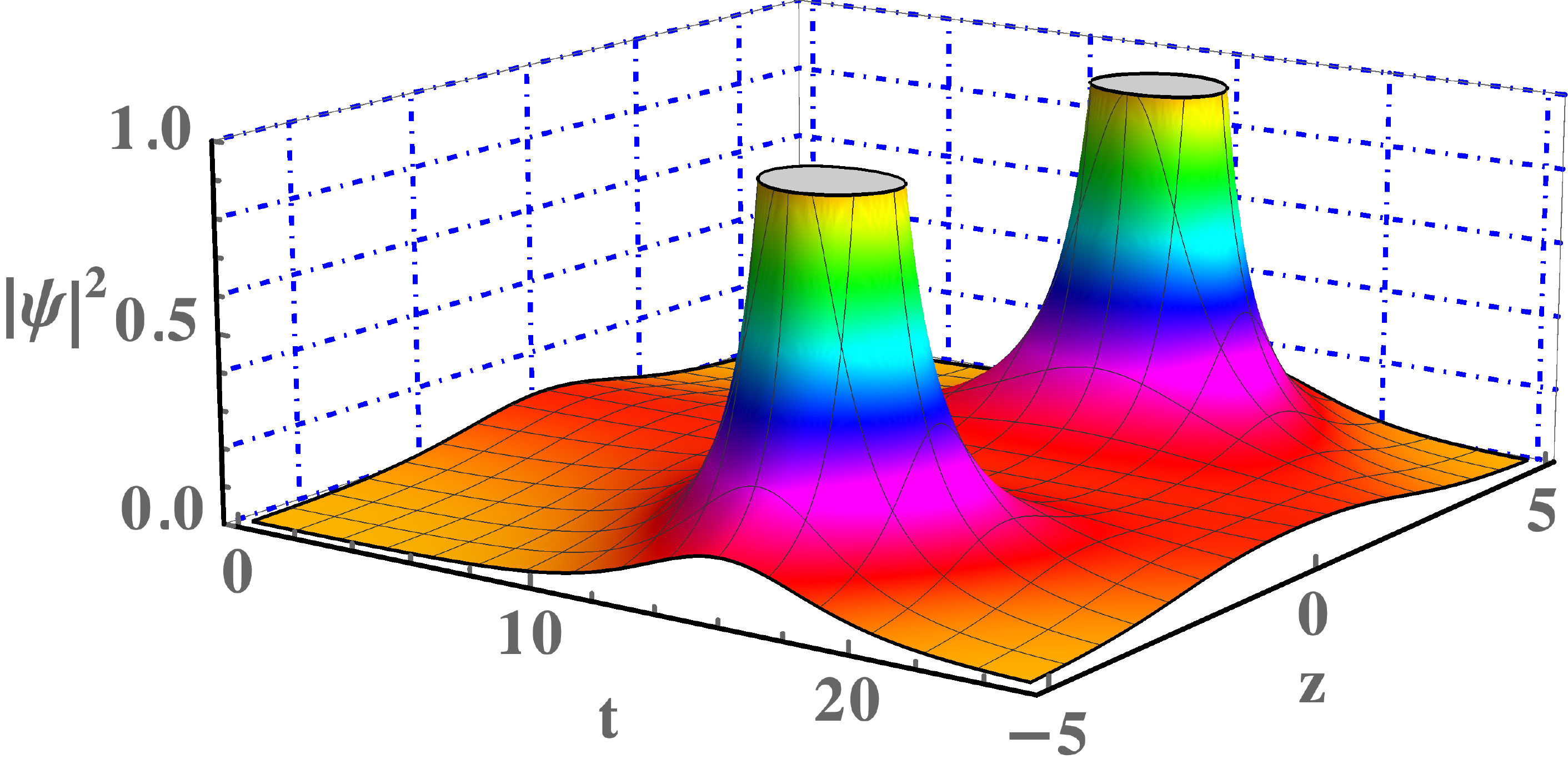}
  \caption{Soliton solution for the unmodulated trap with $M_0=0.01$}
  \label{unmod0.01}
\end{figure}

For constant value of $M_0$ Eq. \eqref{e6} takes the form a free particle Schr\"{o}dinger equation whose eigenfunction is $\phi(t)=\exp(\sqrt{M_0}\ t)$. The control parameters for different values of $M_0$ are obtained by substituting $\phi(t)=\exp(\sqrt{M_0}\ t)$ in Eq. \eqref{e7}. These control parameters affect the spatio-temporal dynamics of the bright solitons which are revealed in Figs. (\ref{unmod0.001}) and (\ref{unmod0.01}). 

It can be seen in Fig. (\ref{unmod0.001}), the soliton, corresponding to $M_0=0.001$ is non-singular. For $M_0=0.01$ the soliton becomes singular as it evolves as shown in Fig. (\ref{unmod0.01}). As $t$ increases, the soliton localizes further gaining in amplitude. A possible explanation for the above singular soliton behaviour is as follows. Traps with bigger height have more energetic condensate atoms as compared to traps with lower height. Hence the dynamics of the soliton in traps with more height pick up in amplitude resulting in a singular soliton. Additionally, we emphasize that the self-focusing effect is responsible for this singularity \cite{weinstein,gaeta}.


\subsection{Temporal modulations using regular and rational oscillator potential functions} 

Here we introduce temporal trap modulations by considering  $V(t)$ in the form of the regular and rational oscillator. The regular oscillator type temporal modulations are of the form
\begin{equation}\label{e8}
    V_{\mathrm{reg}}(t)=t^2-1    
\end{equation}
while the rational oscillator type modulation is as follows
\begin{equation}\label{e9}
 V_{\mathrm{rat}}(t)= t^2+\frac{16(4t^2-2)}{(4t^2+2)^2} - 2.   \end{equation}
The subscripts reg and rat will be used to indicate functions related to the regular and the rational type modulations. It can be clearly seen from the above equations that $ V_{\mathrm{rat}}(t) = V_{\mathrm{reg}}(t) + \mathrm{rational \ terms}$.

\begin{figure}
\centering
  \includegraphics[width=0.9\linewidth]{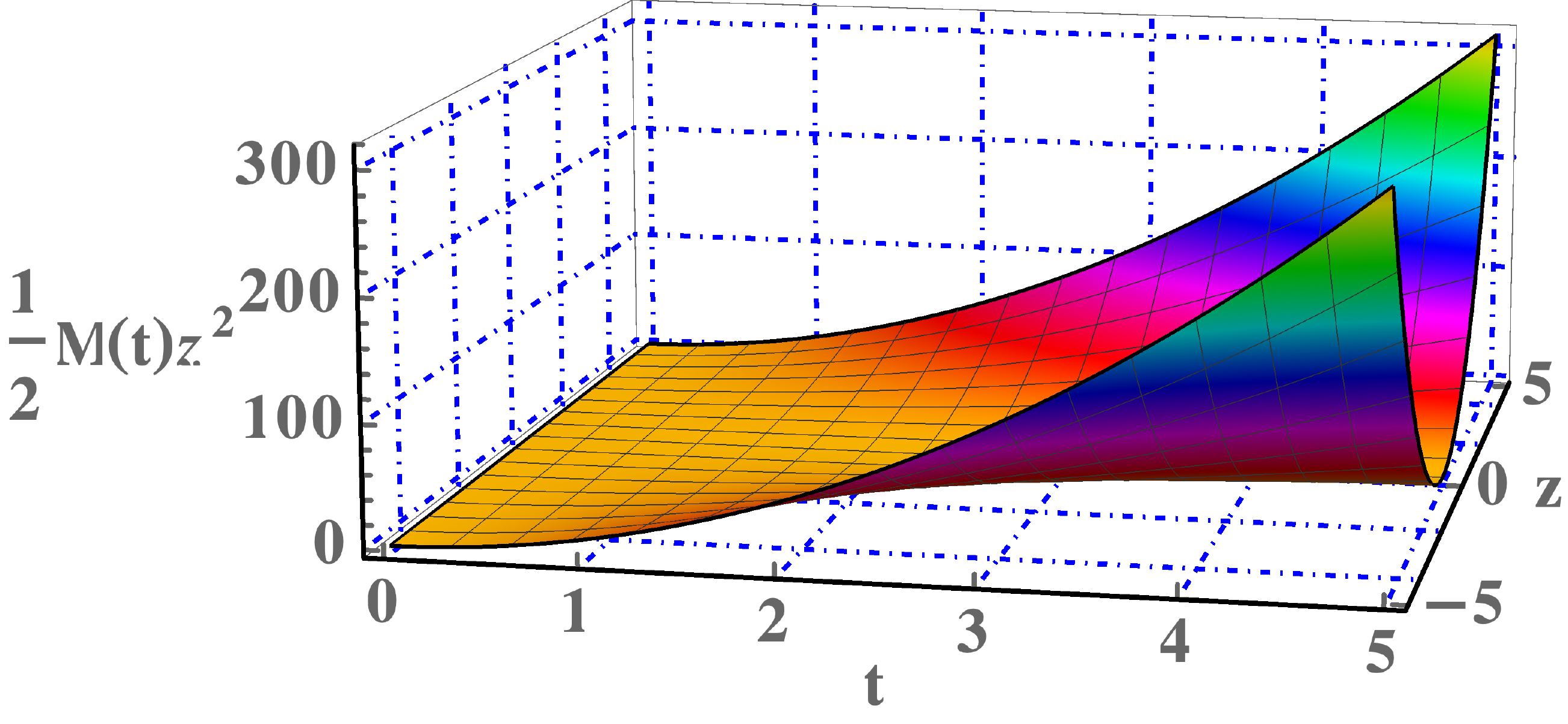}
  \caption{Trap with regular oscillator modulation.}
  \label{reg-osctr}
\end{figure}

\begin{figure}
  \centering
  \includegraphics[width=0.9\linewidth]{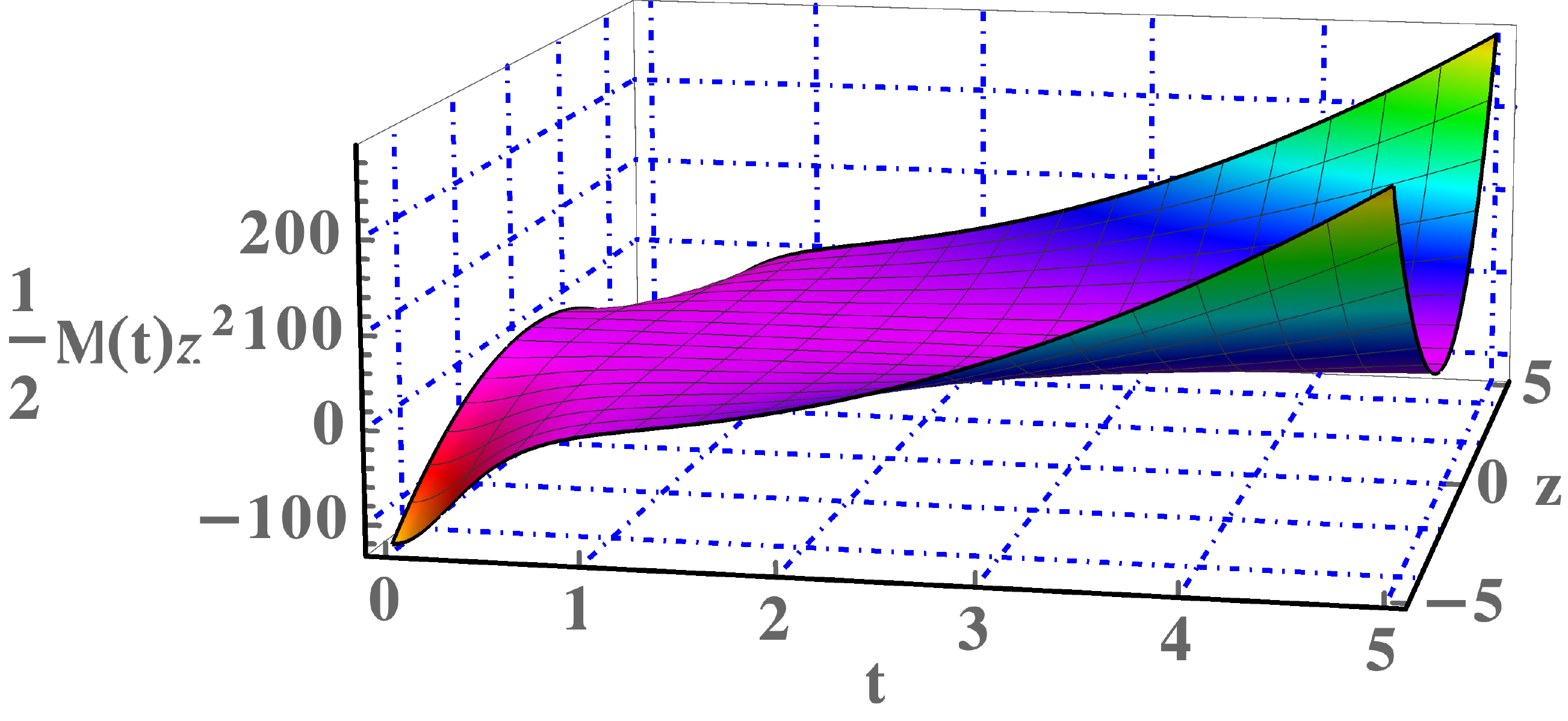}
  \caption{Trap with rational oscillator modulation.}
  \label{rt-osctr}
\end{figure}

The corresponding $\phi(t)$ obtained from \eqref{e6} for $ V_{\mathrm{reg}}(t)$ and $V_{\mathrm{rat}}(t)$ are
\begin{equation}\label{e10}
\phi_{\mathrm{reg}}(t) =\sqrt{\frac{1}{\sqrt{\pi}}} \exp{(-t^2/2)}
\end{equation}
and 
\begin{equation}\label{e11}
\phi_{\mathrm{rat}}(t) =\sqrt{\frac{8}{\sqrt{\pi}}} \frac{\exp{(-t^2/2)}}{4t^2+2}
\end{equation}
respectively. The phase-front curvatures $c(t)$ for the above modulations turn out to be 
\begin{equation}\label{osci-}
    c_{\mathrm{reg}}(t)=t   \,\,\, \mathrm{and} \,\,\, c_{\mathrm{rat}}(t)=t+\frac{8t}{4t^2+2}.
\end{equation}
The corresponding modulated traps for $V_{\mathrm{reg}}$ and $V_{\mathrm{rat}}$ are depicted in the figures (\ref{reg-osctr}) and (\ref{rt-osctr}) respectively. 

\begin{figure}
  \centering
  \includegraphics[width=0.9\linewidth]{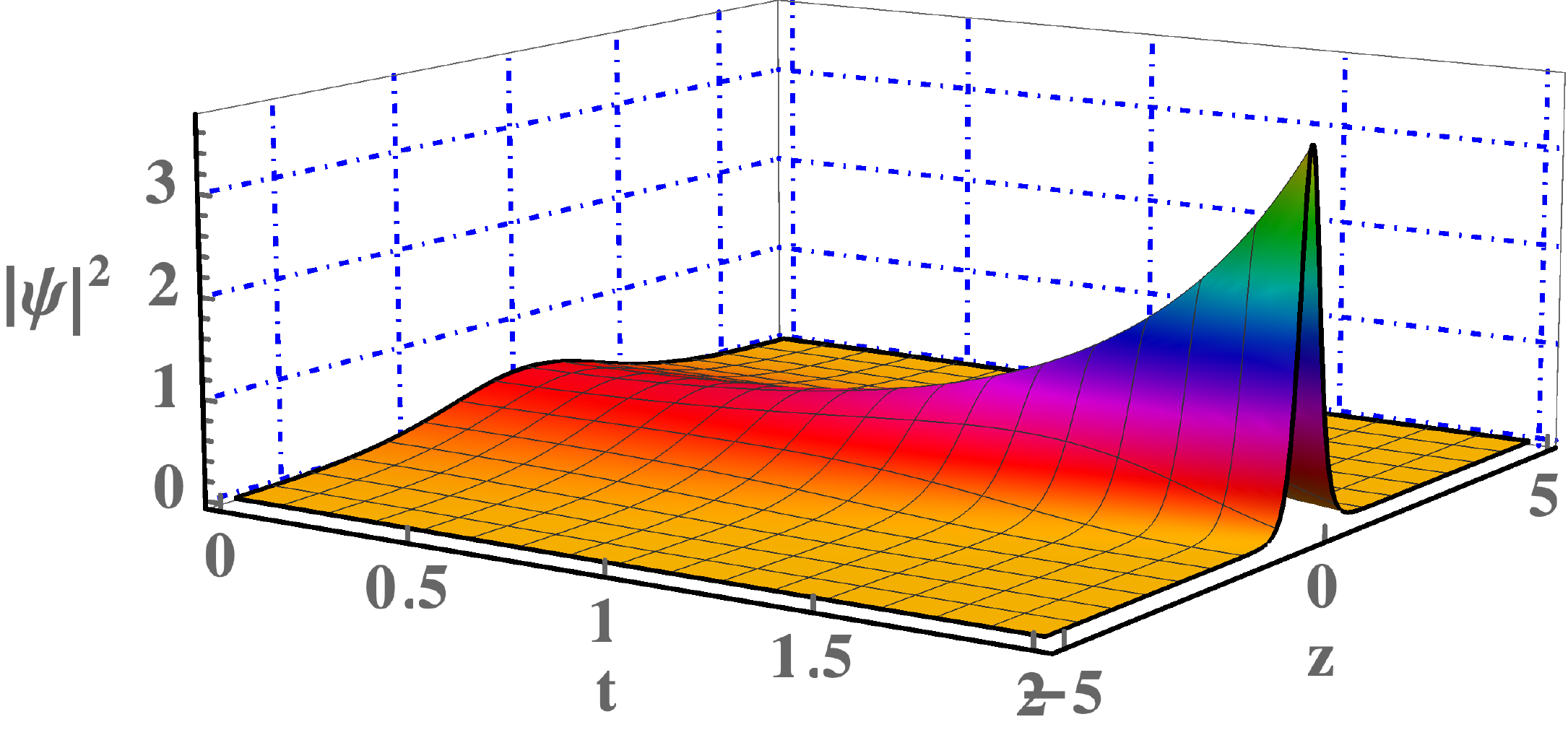}
  \caption{Regular oscillator modulation. Parameter values used are $A_0 = 0.5$, $\gamma_0 = -0.5$, $\ell_0=0$}
  \label{reg-osc1}
\end{figure}

\begin{figure}
  \centering
  \includegraphics[width=0.9\linewidth]{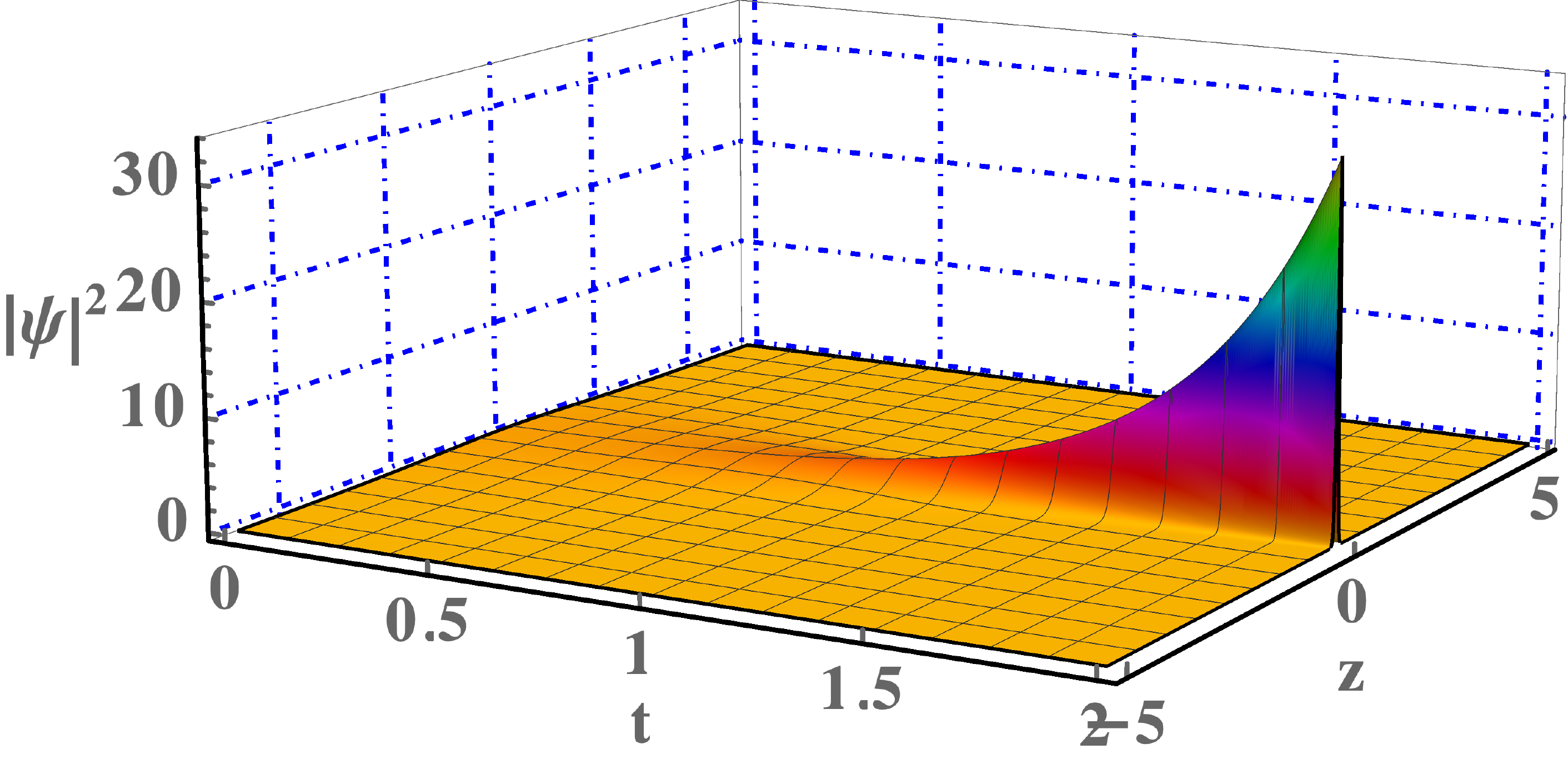}
  \caption{Rational oscillator modulation. Parameter values used are $A_0 = 0.5$, $\gamma_0 = -0.5$, $\ell_0=0$}
  \label{rt-osc1}
\end{figure}

\begin{figure}
\centering
  \includegraphics[width=0.9\linewidth]{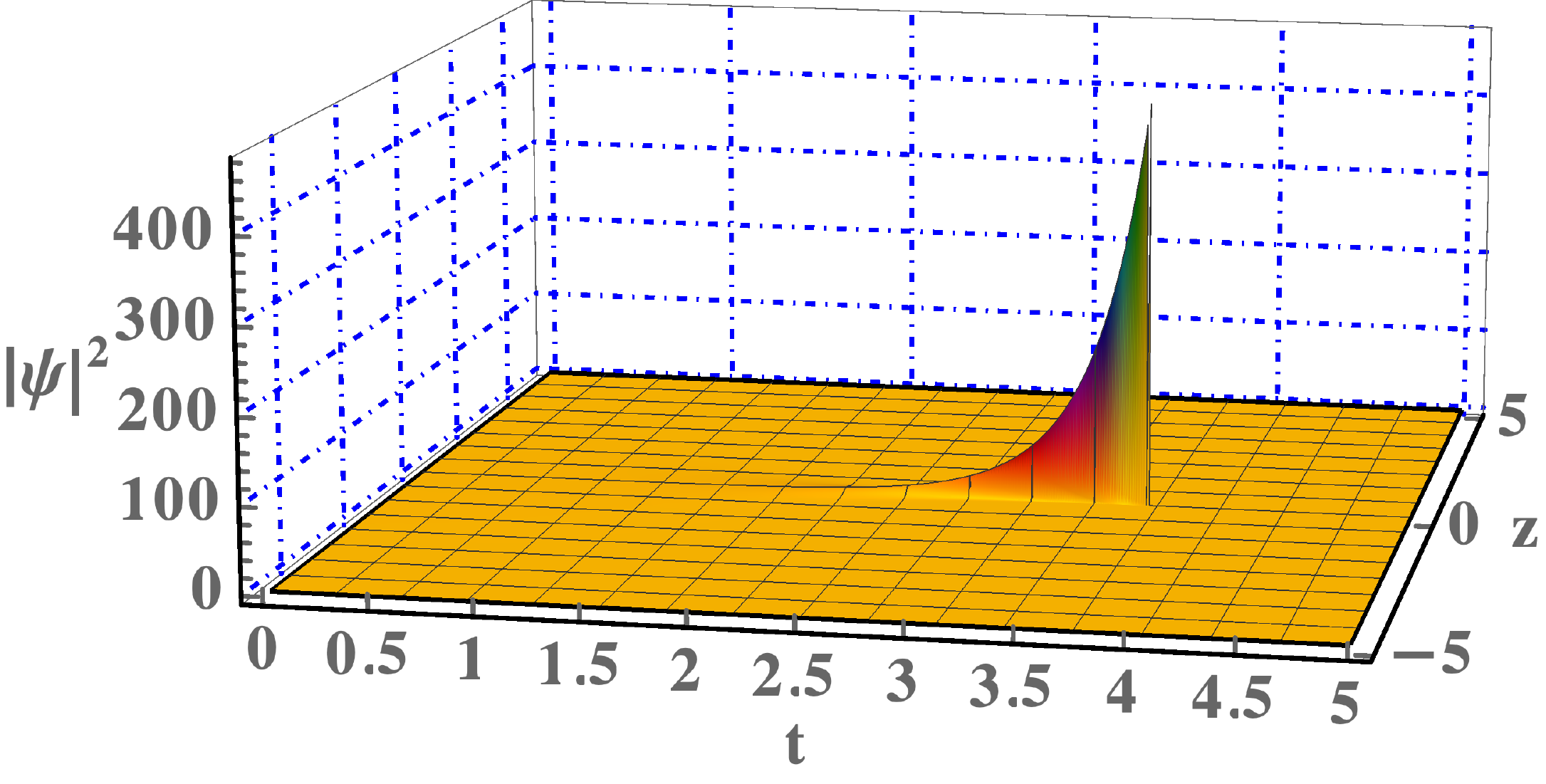}
  \caption{Regular oscillator modulation.Parameter values used are $A_0 = 0.5$, $\gamma_0 = -0.5$, $\ell_0=0$}
  \label{reg-osc2}
\end{figure}

\begin{figure}
  \centering
  \includegraphics[width=0.9\linewidth]{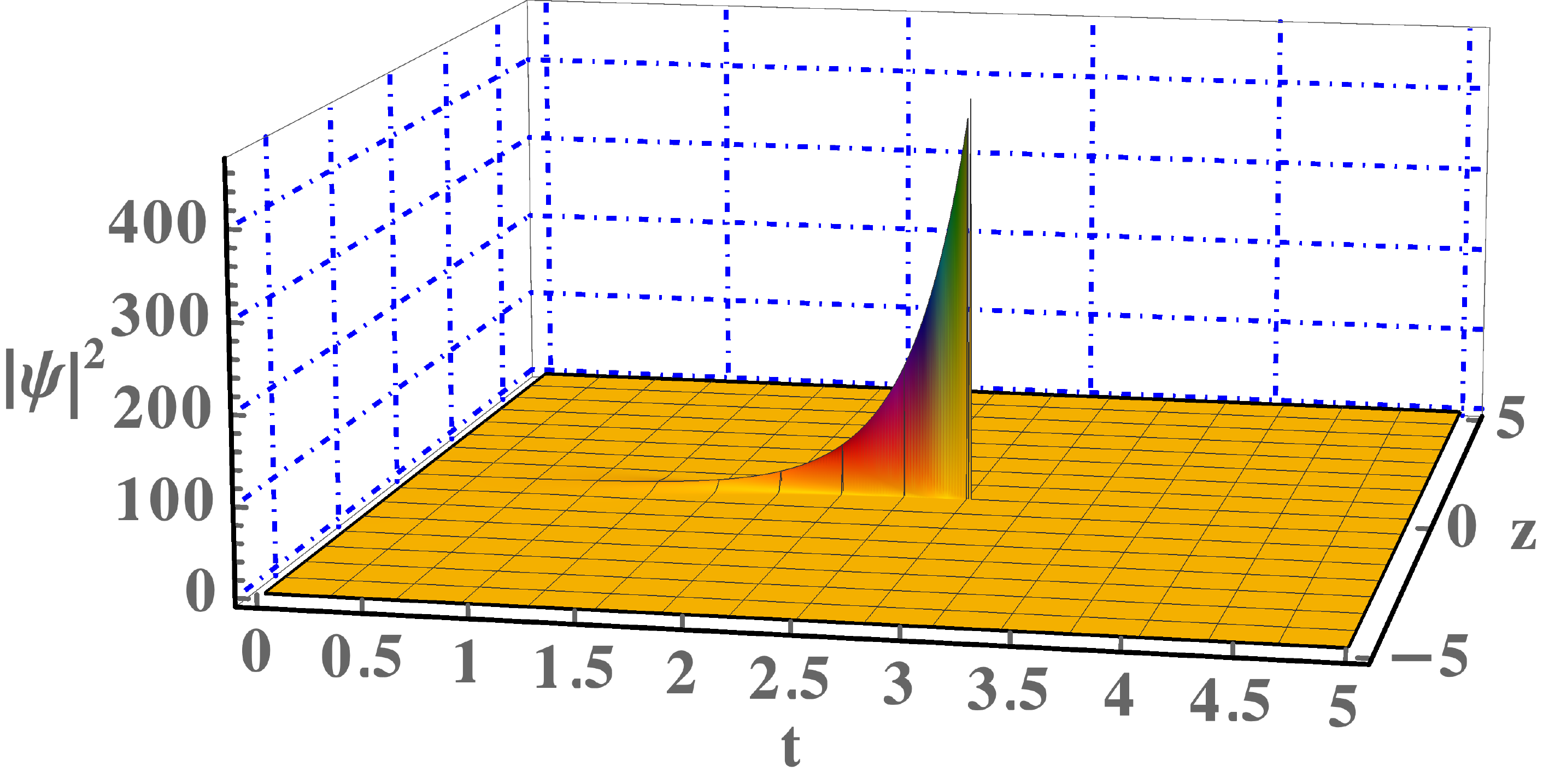}
  \caption{Rational oscillator modulation.Parameter values used are $A_0 = 0.5$, $\gamma_0 = -0.5$, $\ell_0=0$}
  \label{rt-osc2}
\end{figure}

\begin{figure}
\centering
\includegraphics[width=0.9\linewidth]{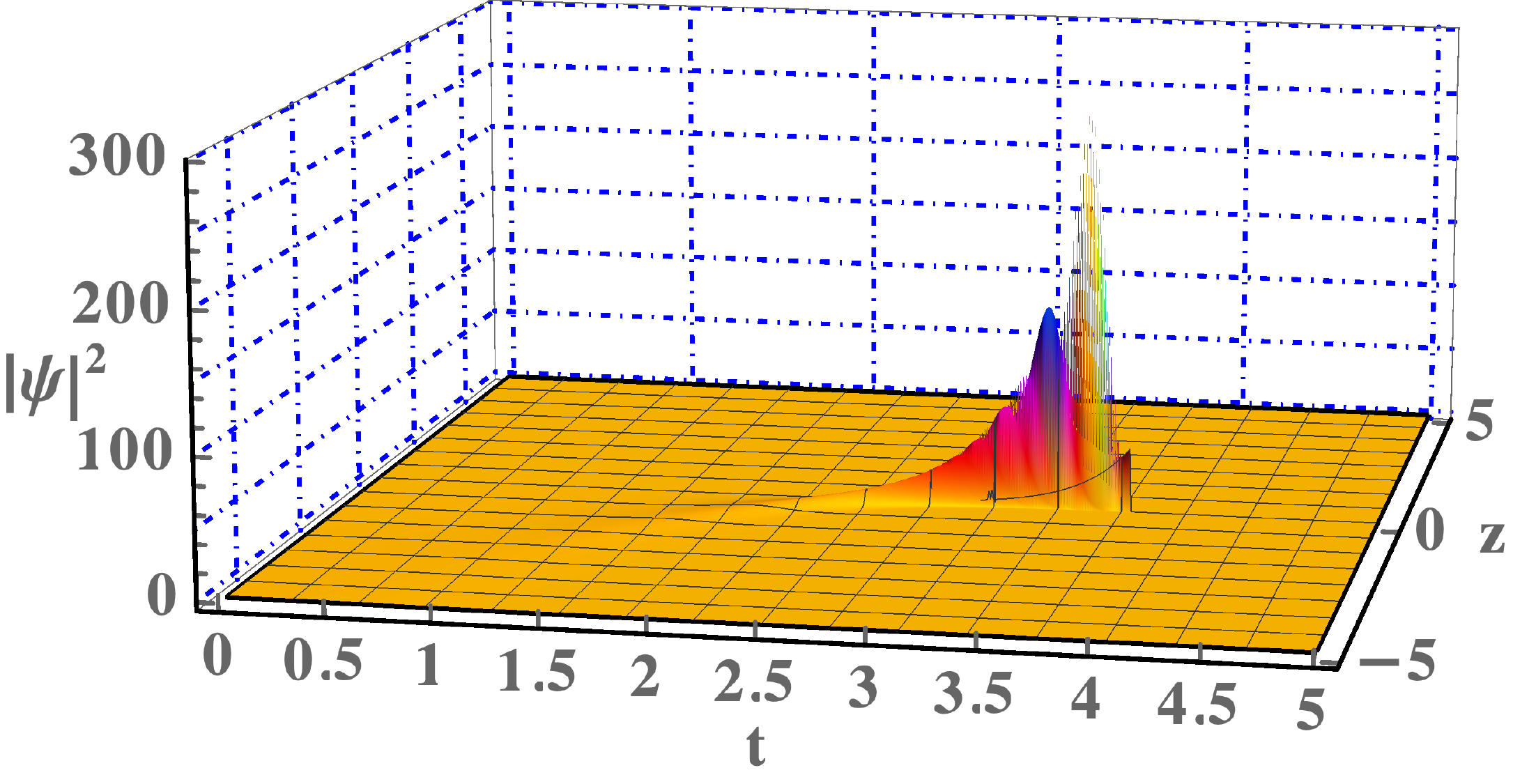}
\caption{Regular oscillator modulation. Parameter values used are $A_0 = 0.5$, $\gamma_0 = -0.5$, $\ell_0=-4$}
\label{reg-osc3}
\end{figure}

\begin{figure}
\centering
\includegraphics[width=0.9\linewidth]{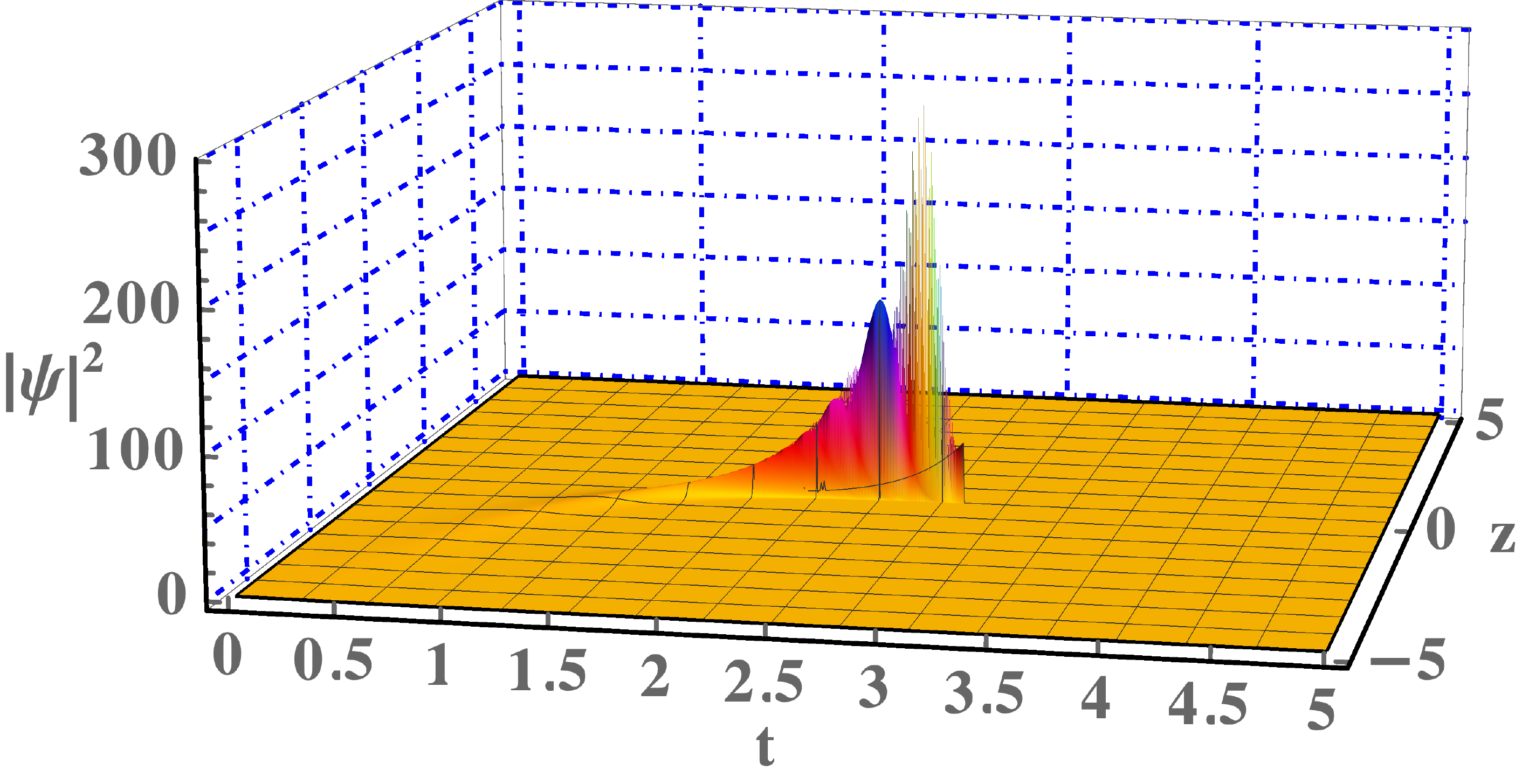}
\caption{Rational oscillator modulation. Parameter values used are $A_0 = 0.5$, $\gamma_0 = -0.5$, $\ell_0=-4$}
\label{rt-osc3}
\end{figure}

The confining traps show initial flattening  with the introduction of these temporal modulations. 
Though the flattening is pronounced initially for both the cases, in the rational case, we see that the presence of the rational terms in fact make the trap repulsive initially, which gradually flattens and becomes confining again as $t$ increases. We can also see that as $t$ increases both the traps become indistinguishable.
Substitution of equations \eqref{e10} and \eqref{e11} in Eq. \eqref{e7} for these two modulations will introduce the affect of the modulations into the solutions of the NLSE given in equation \eqref{bsol}. The surface plots of the condensate, for the two types of modulations are shown in figures (\ref{reg-osc1}), (\ref{rt-osc1}), (\ref{reg-osc2}), (\ref{rt-osc2}),  (\ref{reg-osc3}), (\ref{rt-osc3}). 

We observe that both these temporal modulations lead to the condensate undergoing rapid compression as it propagates through the atomic wave guide. This is clearly visible in the short term behaviour of the condensate in the Fig. (\ref{reg-osc1}) and Fig. (\ref{rt-osc1}). Figures (\ref{reg-osc2}) and (\ref{rt-osc2}) depict the long term behaviour for the same parameter values. It can be seen that the compression rate in the case of $V_{rat}(t)$ is faster. In figures (\ref{reg-osc3}) and (\ref{rt-osc3}), the soliton dynamics is captured when the center of mass parameter is nonzero. It can be seen that this leads to the soliton evolving off the symmetry axis of the trap. In addition $\ell_0$ being nonzero impacts the soliton amplitude. 

\subsection{Modulation using Regular and Rational Scarf I potentials}

Another set of temporal modulations can be produced using the regular and the rational Scarf-I potentials which are of the form 
\begin{equation}\label{sc1reg}
V_{reg}(t)= [\alpha(\alpha-1)+\beta^2] \sec^2 t - \beta(2\alpha-1)\sec t \tan t 
\end{equation}
\begin{eqnarray} \nonumber
V_{rat}(t) & = & [\alpha(\alpha-1)+\beta^2] \sec^2 t - \beta(2\alpha-1)\sec t \tan t \\  \nonumber \\  \label{sc1rat} & + & \frac{2(2\alpha-1)}{2\alpha-1-2\beta\sin t} - \frac{2[(2\alpha-1)^2-4\beta^2]}{(2\alpha-1-2\beta\sin t)^2}
\end{eqnarray}
Here, $\beta<(\alpha-1)$. The traps modulated using equations \eqref{sc1reg} and \eqref{sc1rat} are given in figures  respectively. These two modulations are doubly periodic in time, which deliver kicks of varying intensity at different times to the condensate.  

\begin{figure}
\centering
\includegraphics[width=0.9\linewidth]{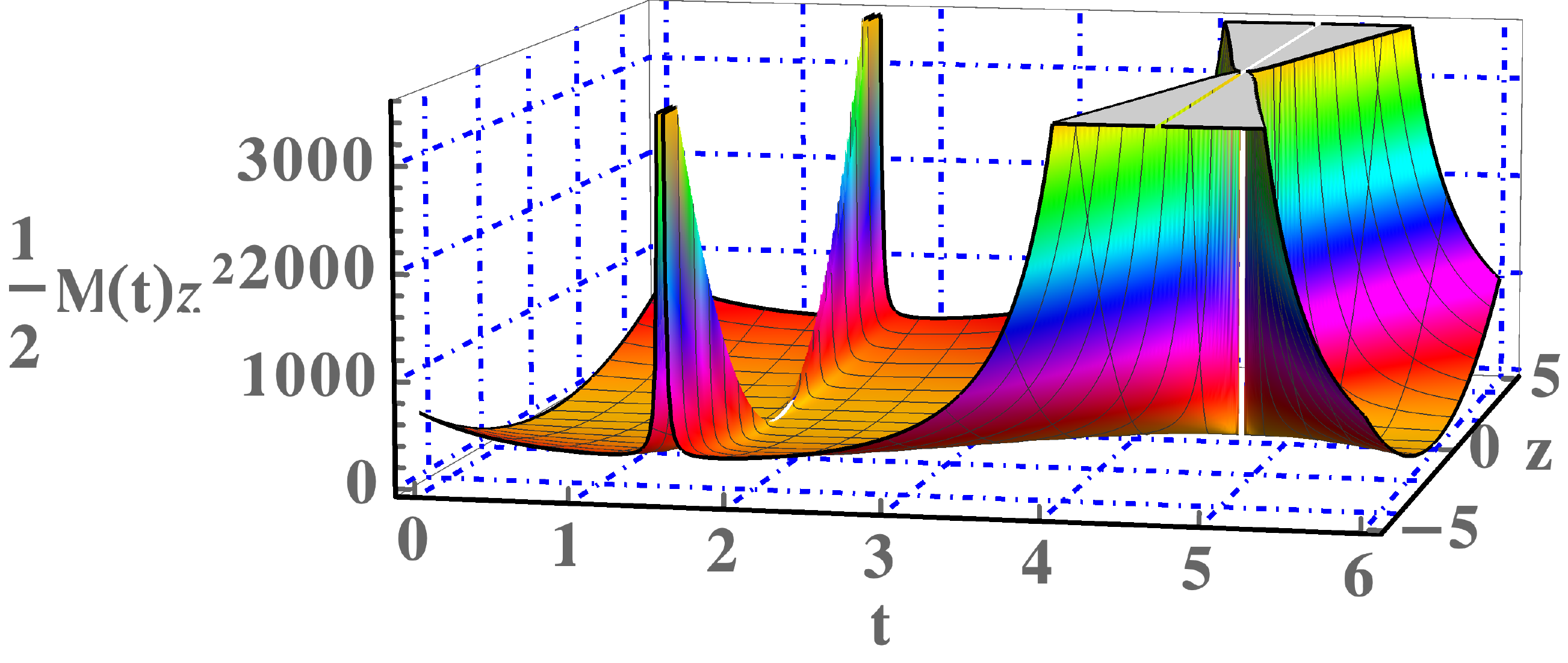}
\caption{Trap with regular Scarf-I modulation $\alpha=6$, $\beta=4.9$.}
\label{reg-scarf1tr}
\end{figure}

\begin{figure}
\centering
\includegraphics[width=0.9\linewidth]{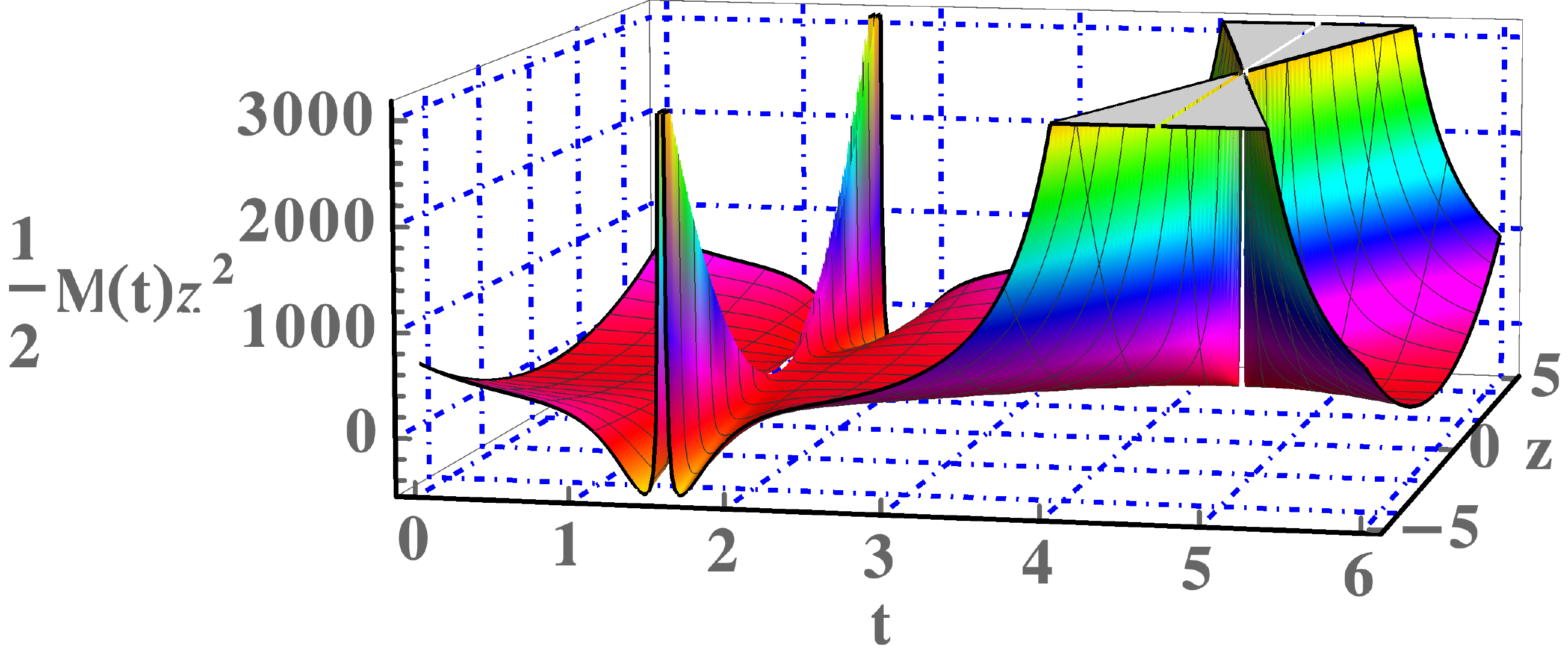}
\caption{Trap with rational Scarf-I modulation $\alpha=6$, $\beta=4.9$.}
\label{rt-scarf1tr}
\end{figure}

From the figures (\ref{reg-scarf1tr}) and (\ref{rt-scarf1tr}) is can be seen that the rational modulation is distinctly different from the regular one for small $t$.

As in the preceding case $\phi(t)$ for the above modulations  are 
\begin{equation}
\phi_{reg}(t) = (1 - \sin t)^{(\alpha - \beta)/2} (1 + \sin t)^{(\alpha + \beta)/2}
\end{equation}
and 
\begin{eqnarray} 
\phi_{rat}(t) & = & \frac{(1 - \sin t)^{(\alpha - \beta)/2} (1 + \sin t)^{(\alpha + \beta)/2}}{2 \alpha - 1 - 2 \beta \sin t} \\ \nonumber
&& \frac{1}{2 (\beta_1 - \alpha_1)} \big[\alpha_1 + \beta_1 + 2 - (\beta_1 - \alpha_1) \sin t \big].
\end{eqnarray}
In the above $\alpha_1 = \alpha - \beta - 1/2$ and $\beta_1 = \alpha + \beta - 1/2$.  For the sake of completeness we also give the the corresponding phase front curvatures for the above two modulations
\begin{equation}
c_{reg}(t) = \alpha \tan t - \beta \sec t
\end{equation}
and
\begin{eqnarray} 
c_{rat}(t) & = & \alpha \tan t - \beta \sec t \\ \nonumber
& - & \frac{2\beta\cos t}{2\alpha-1-2\beta \sin t} + \frac{2\beta\cos t}{2\alpha+1-2\beta \sin t}
\end{eqnarray}
\begin{figure}
\centering
\includegraphics[width=0.9\linewidth]{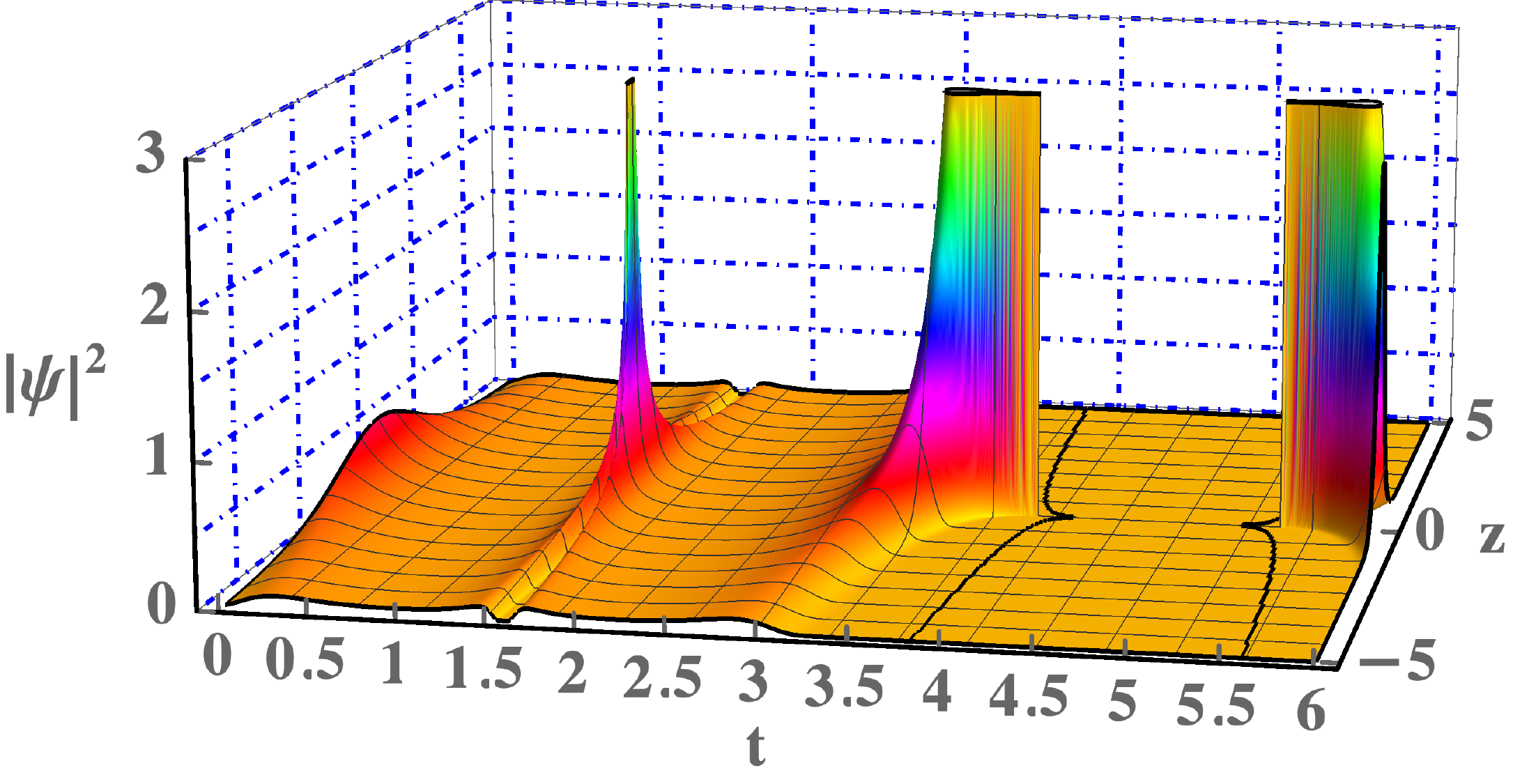}
\caption{Regular Scarf-I modulation.Parameter values used are $A_0 = 0.5$, $\gamma_0 = -0.5$, $\ell_0=0$, $\alpha=6$, $\beta=4.9$}
\label{lt-regscarf1}
\end{figure}
\begin{figure}
\centering
\includegraphics[width=0.9\linewidth]{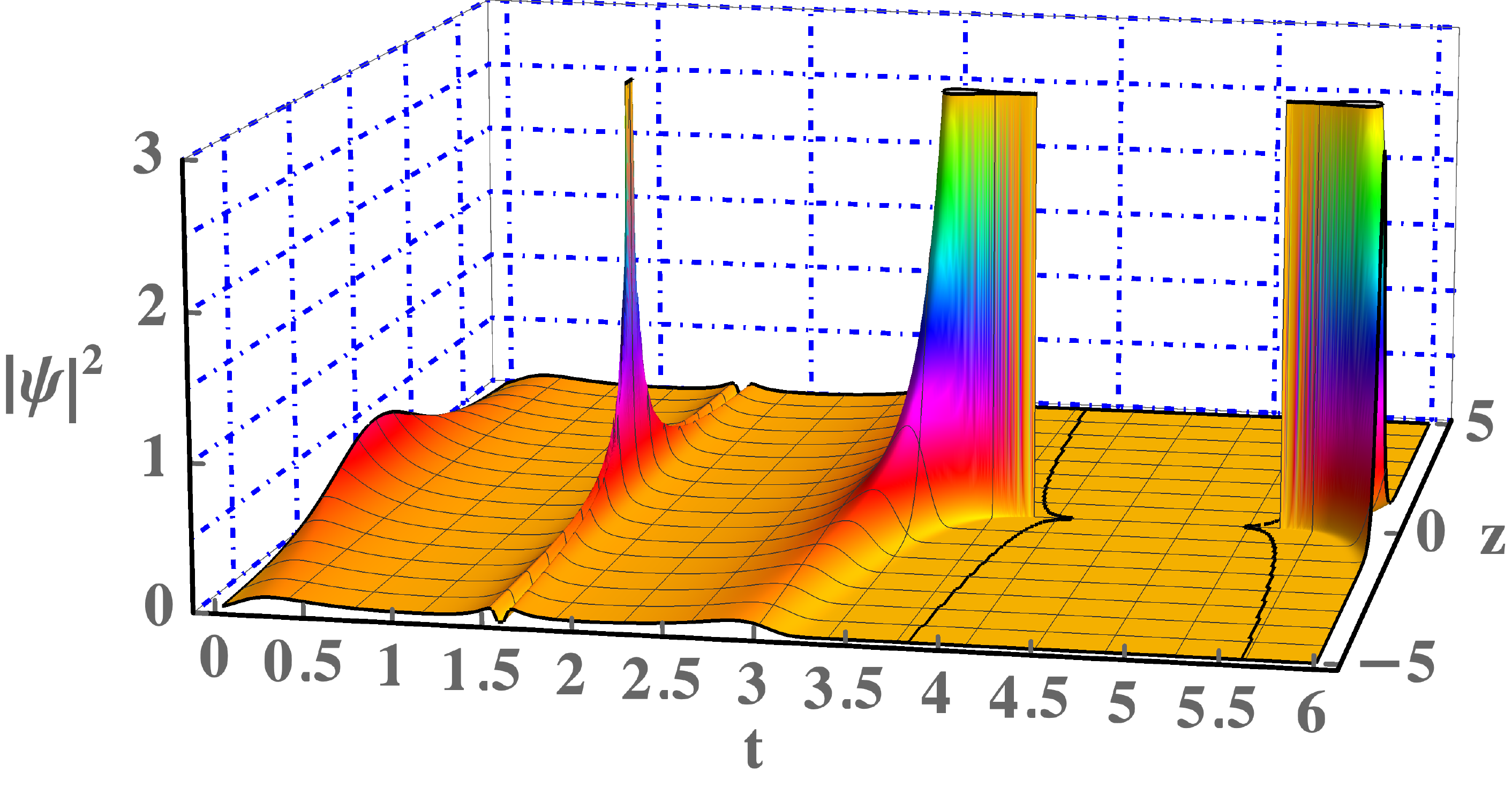}
\caption{Rational Scarf-I modulation.Parameter values used are $A_0 = 0.5$, $\gamma_0 = -0.5$, $\ell_0=0$, $\alpha=6$, $\beta=4.9$}
\label{lt-rtscarf1}
\end{figure}
\begin{figure}
\centering
\includegraphics[width=0.9\linewidth]{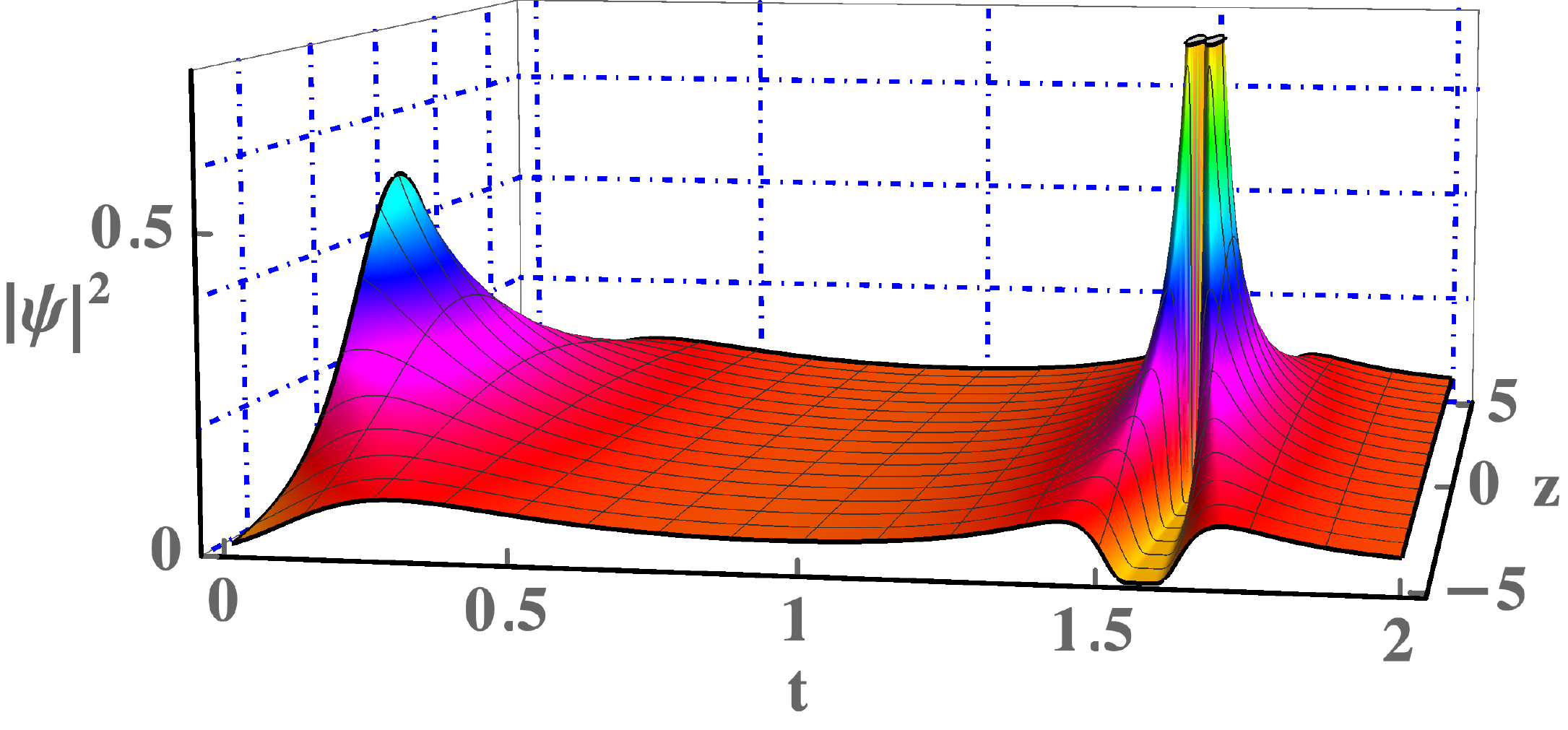}
\caption{Regular Scarf-I modulation. Parameter values used are $A_0 = 0.5$, $\gamma_0 = -0.5$, $\ell_0=0$, $\alpha=6$, $\beta=4.9$}
\label{st-regscarf1}
\end{figure}
\begin{figure}
\centering
\includegraphics[width=0.9\linewidth]{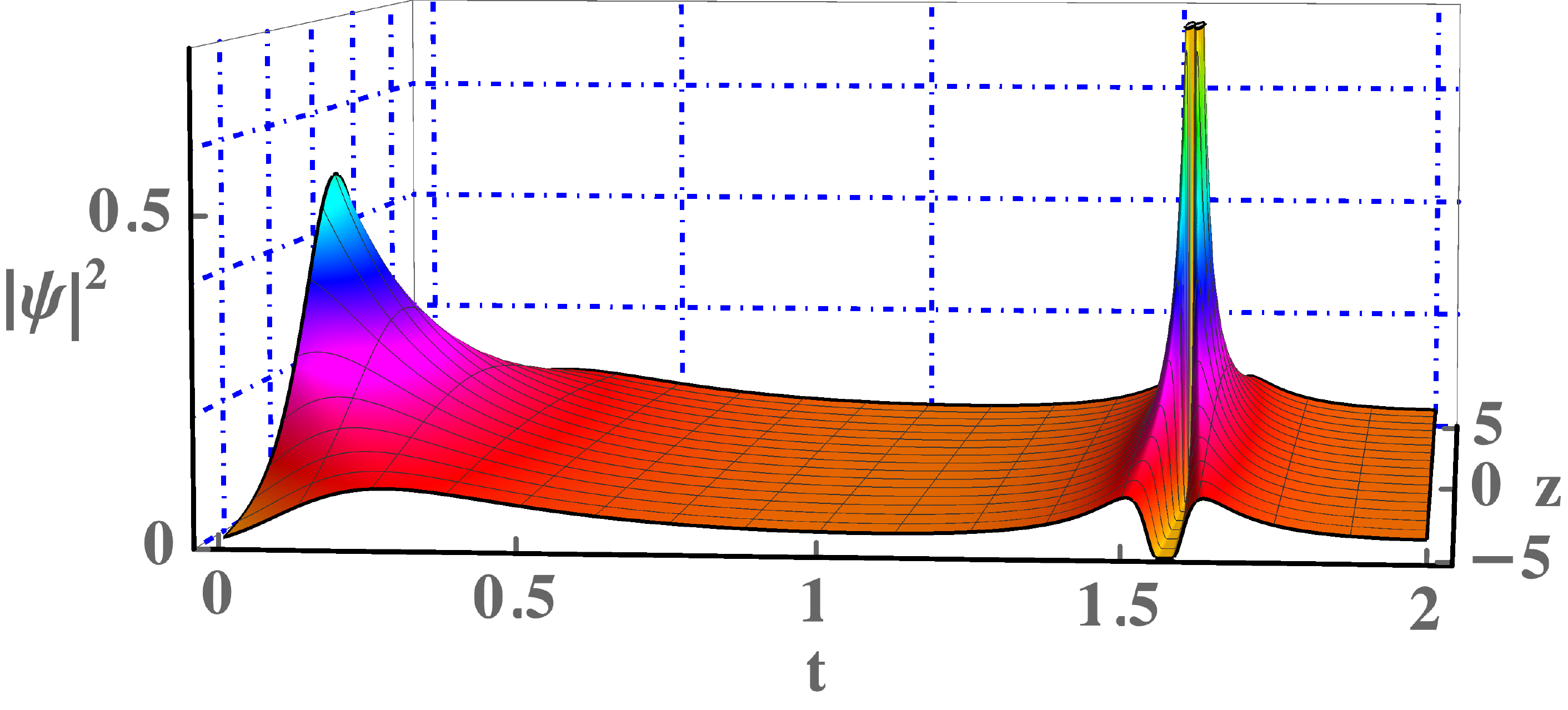}
\caption{Rational Scarf-I modulation. Parameter values used are $A_0 = 0.5$, $\gamma_0 = -0.5$, $\ell_0=0$, $\alpha=6$, $\beta=4.9$}
\label{st-rtscar1}
\end{figure}
From figures (\ref{lt-regscarf1}) and (\ref{lt-rtscarf1}) we can see that the long term behaviour of the soliton is same for both the modulations. The affect of the rational terms can be seen in figure (\ref{st-rtscar1}) around $t=1.5$ where the soliton compression is slightly faster that the regular modulation as shown in Fig. (\ref{st-regscarf1}). This behaviour coincides with the behaviour of the trap modulation around $t=1.5$ in figure (\ref{rt-scarf1tr}).

\section{Conclusions}

In conclusion, we have studied the controllable behavior of nonautonomous matter waves in different "smart" transient trap variations in the context of the cigar-shaped BECs. To accomplish this task, use was made of a novel self-similarity transformation by which we have reduced the nonautonomous GP equation to the elliptic equation that admits soliton solutions. This procedure led to a consistency equation which is in the form of Riccati equation. The connection between the Riccati and the linear Schr\"odinger equation through the Cole-Hopf transformation has been successfully used to introduce temporal trap variations. For our study, we ave explored the one dimensional ES potentials and their newly constructed rational extensions, as functions of time, to introduce interesting temporal trap modulations. As an application, we delineated the rapid pulse compression experienced by the solition due to the rational modulations in the form of rational oscillator and rational Scarf-I.  It is observed that the modulations induced by the rational terms have a definite effect only for small values of $t$, nevertheless the amplitude of the soliton remains the same for rational as well as regular modulations for large $t$. We envision that these newly introduced trap variations may be experimentally realized in the case of cigar-shaped BEC.

\vskip0.5cm

\noindent
{\bf Acknowledgments}\\
TS thanks SERB, India for financial support via Grant: ECR/2015/000081. SSR acknowledges financial support from SERB, India via Grant: EMR/2016/005002.

\end{document}